\documentclass[11pt]{article}
\usepackage{amsmath,amssymb}
\usepackage{setspace,subfig}
\usepackage[hidelinks]{hyperref}
\usepackage{color}
\usepackage{pgf}
\usepackage{setspace}
\usepackage{tikz}
\usepackage{times}
\usetikzlibrary{arrows,shapes.arrows,shapes.geometric,
	shapes.multipart,backgrounds,decorations.pathmorphing,positioning,fit,automata}
\usepackage[latin1]{inputenc}

\usepackage{chngcntr}
\usepackage{amsthm}
\usepackage{enumitem}
\usepackage{amssymb}
\usepackage{natbib}

\usepackage{caption} 

\usepackage{graphicx}

\newtheoremstyle{note}% <name>
{8pt}% <Space above>
{8pt}% <Space below>
{}% <Body font>
{}% <Indent amount>
{\bfseries}% <Theorem head font>
{:}% <Punctuation after theorem head>
{.5em}% <Space after theorem headi>
{}% <Theorem head spec (can be left empty, meaning `normal')>

\theoremstyle{note}

\allowdisplaybreaks

\usepackage[left=1.5in,right=1.5in,top=1in,bottom=1in]{geometry}
\usepackage{algorithm}
\date{}
\usepackage{tikz}

\def\bSig\mathbf{\Sigma}

\newcommand*{\defeq}{\mathrel{\vcenter{\baselineskip0.5ex \lineskiplimit0pt
			\hbox{\scriptsize.}\hbox{\scriptsize.}}}%
	=}

\usepackage[figuresright]{rotating}

\begin{document}

	\title{On the homogeneity of  measures for binary associations}
	\author{Linbo Wang}
	\clearpage \maketitle

\begin{abstract}
  Applied researchers often claim that the risk difference is more heterogeneous than the relative risk and the odds ratio. Some also argue that there are theoretical grounds for why this claim is true. In this note, we point out that these arguments reflect that certain effect measures are variation independent of a nuisance parameter that is easier to interpret, rather than the homogeneity of these measures.
\end{abstract}

\section{Introduction}

  Applied researchers often claim that the risk difference (RD) is a more heterogeneous effect measure than the relative risk (RR) and odds ratio (OR); see \cite{poole2015risk} and references therein. Based on surveys of meta-analyses, they found \emph{empirical evidence} that the null hypotheses of homogeneity are rejected more often for the RD than for the RR and OR. However, some epidemiologists have pointed out that the current empirical evidence for the claim that the RD is more heterogeneous is not satisfactory. Specifically, \cite{poole2015risk} point out that the current empirical results on RD being more heterogeneous may be due to difference in statistical power of tests used. From this perspective, the high power for testing the null of RD or RR may in fact be used as an argument to support the use of RD or RR over OR.
  
  On the other hand, some argue that there are \emph{theoretical grounds} for this claim. For example, \cite{schmidt2016comments} point out that if $p_{00}=0.27$, $p_{10}=0.46, p_{01}>0.81,$ then the RD is not possible to be homogeneous; here $p_{va} = P(Y(V=v, A=a))$ denotes the risk among treatment group $a$ and covariate group $v$. See  \cite{omalley2022discussion} for a similar example. \cite{ding2015differential} theoretically quantify the homogeneity of different effect scales. Under the no-interaction condition for any effect measure, the four outcome probabilities $p_{va}, v=0,1,a=0,1$ lie in a three dimensional space, called the the homogeneity space.  \cite{ding2015differential} compare the three-dimensional volume of the homogeneity space in $\mathbb{R}^4$ and found that the homogeneity space of the RD has the smallest volume, and that of the OR has the largest volume.  

In this note, we argue that the current theoretical arguments for the heterogeneity of the RD is not satisfactory either. These arguments reflect that certain effect measures are variation independent of a nuisance parameter that is easier to interpret (i.e. the baseline risk), rather than the homogeneity of these measures. Note that former is decided by humans, but the latter is decided by nature. We also point out that the volume of the homogeneity space depends on the coordinate system one chooses. Although \cite{ding2015differential}'s calculation holds under the probability scales, one can construct alternative scales under which the volume of the homogeneity space is larger for the RR than the OR. 
Unless we know which coordinate system the nature prefers, it  seems very difficult to argue
which null hypothesis of homogeneity is more likely to hold. Alternatively, one can say under the prior information of a specific coordinate system, one effect measure is more heterogeneous than another.

\section{Background and notation}
	
Let $A = 0,1$ be a binary exposure, $Y=0,1$ be a binary response, and $V=0,1$ be a  binary covariate. Define $p_{av} = P(Y\mid A=a, V=v)$ and $p_a(V) = P(Y\mid A=a, V).$ We shall focus on the following effect measures in this article: $RR(v) = p_{1v}/p_{0v}, RD(v) = p_{1v}-p_{0v}, OR(v) = p_{1v}(1-p_{0v})/\{p_{0v}(1-p_{1v})\}.$
% We assume $A$ is randomized.

Unlike the OR, both RD and RR are \emph{variation dependent} on the baseline risk $p_0(V)$. For example, if $p_0(v) = 0.5$, then the range of possible values for the $RD(v)$ is restricted to be $[0.5,-0.5],$ while that for the $RR(v)$ is restricted to be $[0,2].$ In contrast, under the same condition, the range for the $OR(v)$ coincides with its range without this condition, i.e. $[0,\infty)$.

Take the RR as an example. First, consider  the saturated Poisson model:
\begin{flalign*}
	\log RR(V) &= \alpha_0^P + \alpha_1^P V; \\
	\log p_0(V)  &= \beta_0^P + \beta_1^P V,
\end{flalign*}
where $
			\beta_0^P = \log p_{00},
			\beta_1^P = \log p_{10} - \log p_{00},
			\alpha_0^P = \log p_{01} - \log p_{00},
			\alpha_1^P = \log p_{11} - \log p_{10} - (\log p_{01} - \log p_{00}).
$ The model on $p_0(V)$ is a nuisance model as it is not of primary interest. However, as pointed out above, it is variation dependent on $RR(V).$

 \cite{richardson2017modeling} discover that the so-called odds ratio $OP(v) = p_{1v}p_{0v}/\{(1-p_{1v})(1-p_{0v})\}$ is \emph{variation independent} of both the RR and RD. In other words, the range of possible values for $RD(v)$ or $RR(v)$ does not depend on the value of $OP(v)$. Based on this, an alternative model for estimating the RR is
\begin{flalign*}
\log RR(V) &= \alpha_0^P + \alpha_1^P V; \\
\log OP(V)  &= \beta_0^R + \beta_1^R V,
\end{flalign*}
where $
\beta_0^P = \log OP_{0\cdot},
\beta_1^P = \log OP_{1\cdot} - \log OP_{0\cdot},
\alpha_0^P = \log p_{01} - \log p_{00},
\alpha_1^P = \log p_{11} - \log p_{10} - (\log p_{01} - \log p_{00}).
$

\section{Volumes of the Homogeneity Spaces}
\cite{ding2015differential} impose a uniform distribution on $(p_{00},p_{10},p_{01})$, which induces a distribution on $(\beta_0^P, \beta_1^P, \alpha_0^P)$. They then calculated the probability that  $(\beta_0^P, \beta_1^P, \alpha_0^P)$ is compatible with the condition $\alpha_1^P=0$ under this induced distribution. 
Because $\alpha_1^P$ is variation independent of $(\beta_0^P, \beta_1^P, \alpha_0^P)$, and hence  $(p_{00},p_{10},p_{01})$, it is not surprising that the probability is smaller than 1. In fact, according to \cite{ding2015differential}, this probability is 0.75.

\cite{ding2015differential} conducted a similar calculation for the OR. Following a similar argument, since $\alpha_1^L \defeq \log odds_{11} - \log odds_{10} - (\log odds_{01} - \log odds_{00})$ is variation independent of $(p_{00},p_{10},p_{01})$, the corresponding probability is 1.

The key point in \cite{ding2015differential}'s arguments is that there are certain values of $(p_{00},p_{10},p_{01})$ that are not compatible with $\alpha_1^P=0$, but are compatible with $\alpha_1^L=0$. Hence under any $(p_{00},p_{10},p_{01})$, it is more likely that  $\alpha_1^P=0$ is rejected.

However, there is no reason to believe that nature would impose a prior distribution on  $(p_{00},p_{10},p_{01})$, which are variation independent of $\alpha_1^L$ but variation dependent of $\alpha_1^P$. For example, if the nature imposes a prior distribution on $(\alpha_0^P, \beta_0^R, \beta_1^R)$, then the corresponding probability for $RR$ would be 1 since $(\alpha_0^P, \beta_0^R, \beta_1^R)$ is variation independent of $\alpha_1^P$.  Similarly, since $(\alpha_0^P, \beta_0^R,  \beta_1^R)$ is variation independent of $\alpha_1^L,$ the corresponding probability for $OR$ would also be 1.

To further illustrate this point, we consider a nuisance parameter, $$\eta(Y|A,V) \defeq \left|\log \dfrac{(1-p_0(V))(p_1(V)+0.5)}{(1-p_1(V))p_0(V)}\right|.$$
The nuisance parameter is made up so that is variation independent of $RR$, but variation dependent of $OR$.
We note that although this measure seems unnatural for human beings, there is no reason to think this is unnatural for   nature. 

Following the notation above, define the following saturated model
\begin{flalign*}
\log RR(Y|A,V) &= \alpha_0^P + \alpha_1^P V; \\
\log \eta(Y|A,V)  &= \beta_0^\eta + \beta_1^\eta V.
\end{flalign*}

Following  similar arguments as before, if the nature imposes a prior distribution on $(\alpha_0^P, \beta_0^\eta,  \beta_1^\eta)$, then the probability that $(\alpha_0^P, \beta_0^\eta,  \beta_1^\eta)$ is compatible with the condition $\alpha_1^P = 0$ (or equivalently, $RR=1$) would still be 1, but the corresponding probability for $\alpha_1^L = 0$ (or equivalently, $OR=1$) would be smaller than 1!

\section{Concluding remarks}

%  Seems hopeless: what is next?
% 	\begin{itemize}
% 		\item We should not choose measure based on statistical property anyway!
% 		\item Even if we prefer one measure for the sake of generalizability, collapsibility/transportability should be a more important issue: that deals with generalizability of findings within the same data set!
% 		\item This a problem that is theoretically hopeless and practically not important.
% 	\end{itemize}

In this note, we present counterarguments for the claim that  nature would think RD and RR to be more heterogeneous than OR.   We argue that this common perception is due to some parameterizations being more natural to human beings. 

In the literature, it is often claimed  that if a measure is more homogeneous under one scale, then it  would be preferable to use this scale as a measure of the treatment effect for the sake of generalizability \citep[e.g.][]{ding2015differential}. We argue that in some contexts that rely on generalizability such as meta-analysis, collapsibility is also important as it reflects the generalizability of findings within
the same data set. Furthermore, as advocated in \cite{richardson2017modeling}, one should not choose the effect measure based on statistical properties anyway.

%  We also comment that in general, unless in the unusual saturated case, the Poisson/linear model for RR/RD is likely to be wrong. Hence the RR/RD model as described in \cite{richardson2017modeling} is still preferable.

		\bibliographystyle{apalike}
		\bibliography{causal}

\end{document}